\documentclass[twocolumn,showpacs,amsmath,amssymb]{revtex4}
\usepackage{graphicx}
\usepackage{dcolumn}
\usepackage{bm}
\newcommand{\be}{\begin{eqnarray}}
\newcommand{\en}{\end{eqnarray}}
\newcommand{\ben}{\begin{eqnarray*}}
\newcommand{\enn}{\end{eqnarray*}}

\newcommand{\bi}{\begin{itemize}}
\newcommand{\ei}{\end{itemize}}
\newcommand{\im}{\item}

\newcommand{\R}{\Rightarrow}

\begin{document}
\title{A study on bifurcation diagrams in relation to synchronisation in chaotic systems}
\author{Sagar Chakraborty}
\email{sagar@bose.res.in}
\affiliation{S.N. Bose National Centre for Basic Sciences, Saltlake, Kolkata 700098, India}
\author{D. Dutta}
\email{debabrata@bose.res.in}
\affiliation{S.N. Bose National Centre for Basic Sciences, Saltlake, Kolkata 700098, India}
\date{December 10, 2006}
\begin{abstract}
We numerically study some of the 3-D dynamical systems which exhibit complete synchronisation as well as generalised synchronisation (GS) to show that these systems can be conveniently partitioned into equivalent classes facilitating the study of bifurcation diagrams (BDs) within each class.
We demonstrate how BDs may be helpful in assessing the robustness of GS and in predicting the nature of the driven system by knowing the BD of driving system and {\it vice versa}.
We extend the study to include the possible GS between elements of two different equivalent classes by taking the example of the R$\ddot{\textrm{o}}$ssler-driven-Lorenz-system.
\end{abstract}
\pacs{05.45.Xt, 05.45.Gg}
\maketitle
\section{INTRODUCTION}
Ultra-sensitivity on the initial conditions makes it hard to predict the future behaviour of a chaotic system which is termed to have long term unpredictable behaviour.
The question if two chaotic system running side by side may be forced to follow the same (or functionally related) paths on the respective attractors did raise interest in the scientific community long back with the works of Yamada and Fujisaka\cite{1a,1b}; and Afraimovich {\it et. al.}\cite{2} who dealt with coupled chaotic systems.
The subject of synchronisation was thus brought forth.
The theory of synchronisation was given a polished look by Pecora and Carroll\cite{3} who along with the seminal paper by Otts {\it et. al.}\cite{4} have simulated wide range of research activities that has spanned a period of almost two decades.
With possible applications in the fields of electronics, secured communication system, lasers, chemical reactions, biology, robotics etc., synchronisation is, no doubt, a topic of recent interest.
For details see review by Pecora {\it et. al.}\cite{5} and the references therein.
Various possible types of synchronisation are: {\it complete, lag, phase, burst} and {\it generalised} (GS); the last one was introduced for unidirectionally coupled systems in a paper\cite{6} by Rulkov {\it et. al.}
Kocarev and Parlitz\cite{7} went on to develop a general theory for GS and showed that it implied {\it predictability}.
Also, mutual exclusiveness of GS and {\it conditional equivalence} was showcased.
The paper also touched upon the influence of parameter mismatch on GS and hence, {\it robustness} of GS was discussed.
\\
\indent Our interest here is to introduce a systematic study of the relationship between bifurcation diagram (BD) and GS in unidirectional coupled systems (master(drive/driving)-slave(response/driven) systems) in the literature of synchronisation and to discuss its important applications, {\it e.g.}, assessing the robustness of GS etc.
Numerical experimental results will also be presented in due course to validate the claims made in the paper. 
\section{EXTENDED SYNCRONISATION}
Let the following two equations be the driving system and the driven system respectively:
\begin{eqnarray}
\begin{array}{l}
\dot{\bf{x}}={\bf{f}}({\bf{x}};p_x)\\
\dot{\bf{y}}={\bf{g}}({\bf{y}},{\bf{h}}({\bf{x}});p_y)
\end{array}
\label{1}
\end{eqnarray} 
where, ${\bf{x}}\in\mathbb{R}^n$, ${\bf{y}}\in\mathbb{R}^m$, $p_x$ and $p_y$ denote the corresponding sets of parameters, and ${\bf{h}}({\bf{x}})$ is the function responsible for driving the driven system.
Now, for a given value of $p_x$ and $p_y$ GS is said to exist for (\ref{1}) if $\exists$ a transformation ${\bf{H}}_{p_x,p_y}:\mathbb{R}^n\rightarrow\mathbb{R}^m$, a manifold $M=\{({\bf{x}},{\bf{y}})\ni{\bf{y}}={\bf{H}}_{p_x,p_y}({\bf{x}})\}$ and a subset $B=B_x\times B_y\subset\mathbb{R}^n\times\mathbb{R}^m$ with $M\subset B$ $\ni$ all the trajectories of (\ref{1}) with initial conditions $({\bf{x}}_0,{\bf{y}}_0)$ in basin $B$ approach $M$ as $t\rightarrow\infty$.
Obviously, complete or identical synchronisation (IS) means ${\bf{H}}_{p_x,p_y}$ is an identity transformation.
The necessary and sufficient condition\cite{7} for GS to occur is that $\forall ({\bf{x}}_0,{\bf{y}}_0) \in B$, $\dot{\bf{y}}={\bf{g}}({\bf{y}},{\bf{h}}({\bf{x}});p_y)$ is asymptotically stable.
This means one can {\it predict} the behaviour of ${\bf{y}}$, based on the knowledge of ${\bf{x}}$ and ${\bf{H}}_{p_x,p_y}$.
Also, one may define the vector field $\dot{\bf{y}}$ to be {\it conditionally equivalent} to $\dot{\bf{x}}$ if $\exists$ a $C^k$ diffeomorphism which takes orbits of ${\bf{f}}$ to orbits of ${\bf{g}}$, preserving the senses.
Again, suppose that IS occurs for $p_x=p_y=p_0$ (say), and if $\exists$ a neighbourhood $P$ of $p_0 \ni\forall p_y\in P$ driven system is asymptotically stable, one would say that GS persists and hence GS is {\it robust}.
\\
\indent Now we, for the sake of convenience without losing generality, fix all the parameters of driving system and driven system except one for each and let $p_x$ and $p_y$ denote these two.
We keep $p_y$ fixed and vary $p_x$ continuously and suppose that GS persists, unlike IS, for $p_x\in P$.
This independence of GS from $p_x$ implies some sort of robustness which in general need not be same as the one define earlier.
To be more concrete, we illustrate our idea with the following Lorenz system as the driving system:
\be
\dot{x_1}&=&\sigma(x_2-x_1)\nonumber\\
\dot{x_2}&=&rx_1-x_2-x_1x_3\\
\label{2}
\dot{x_3}&=&x_1x_2-bx_3\nonumber
\en
which drives a similar driven system with $y_1$ replaced by $x_1$ and $r$ replaced by $r-\Delta r$.
If one defines ${\bf{e}}={\bf{x}}-{\bf{y}}$, then the evolution of the trajectories in the {\it error space} defined by ${\bf{e}}$ would depend on $\Delta r$.
This means if for $p_x\in P$, such a driving system keeps the driven system in generalised synchrony that becomes IS when $p_x=p_y=p_0\in P$, then effectively we have a family of systems (same coupled differential equations but different parameter value of a certain parameter) such that any two can be in GS.
This can be checked by plotting, for the entire $P$, the maximum {\it conditional Lyapunov exponent} which of course should be negative.
This can be taken as an alternatively as the definition of $P$.
One such plot (fig. (1)) has been illustrated for the R$\ddot{\textrm{o}}$ssler system.
%
\begin{figure}[h!]
\centering
\includegraphics[width=8.0cm]{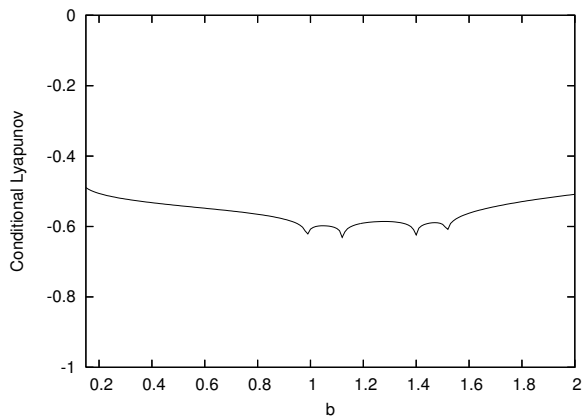}
\caption{ES in unidirectionally coupled R$\ddot{\textrm{o}}$ssler system: maximum conditional Lyapunov exponent has been plotted with the parameter $b$. (One may refer \cite{9} for the standard form of the system.)}
\end{figure}
%
There is no dearth of such families as our numerical experiments with systems like Lorenz, R$\ddot{\textrm{o}}$ssler, Chen, Rucklidge, Nose-Hoover, Rabinovich-Fabrikant equation, Duffing's two well oscillator etc. would suggest.
For such a system, we shall say that the system possess {\it extended synchronisation} (ES) with respect to the parameter $p$ (say).
\\
\indent Let us call such a particular family $F^P$ (where the superscript stands for the neighbourhood of the parameter) and its $i^{\textrm{th}}$ element $f_i$; one may say $f_i$ and $f_j$ are {\it extendedly synchronised} to each other.
One may note that the relation (say, $E$)--``is extendedly synchronised to"-- is an {\it equivalence relation}, for, it satisfies
\begin{enumerate}
\im {Reflexivity}: $f_iEf_i\phantom{x}\forall f_i\in F^P$.
\im {Symmetry}: $f_iEf_j\R f_jEf_i\phantom{x}\forall f_i,f_j\in F^P$.
\im {Transitivity}: $f_iEf_j\phantom{x}\textrm{and}\phantom{x} f_jEf_k$\\$\phantom{xxxxxxxxx}\R f_iEf_k \phantom{x}\forall f_i,f_j,f_k\in F^P$
\end{enumerate}
What this amounts to is also mathematically interesting\cite{8}.
It says that if we form a set $\Phi$ of all such families $F_i^{P_i}$ then each $F_i^{P_i}$ is an equivalence class and forms quotient space.
We shall call such a family {\it extendedly-synchronised-family} (ESF).
Any system $f_i$ in a particular ESF, $F_i^{P_i}\subset \Phi$ would then be the representative of the ESF.
\section{GS, ES AND BD}
Now suppose we pick a particular ESF, F (say) and choose arbitrarily an element of it $f_{driven}$ (say). The question: can we induce bifurcation in the system $f_{driven}$ without changing its parameter which is kept fixed at $p_0$?
The answer is, yes.
We just need to couple $f_{driven}$ with any other element of $F^P$, $f_{driving}$ (say) and continuously vary the parameter, $p$ of  $f_{driving}$ $\ni p\in P$and then plot a variable of $f_{driving}$ {\it vs.} $p$ and the corresponding {\it conjugate variable} of $f_{driven}$ {\it vs.} $p$ to find that both the BDs are almost exactly identical.
See fig.(2) and fig.(3) and their captions for examples validating this point.
\begin{figure}[h!]
\centering
\includegraphics[width=9.0cm]{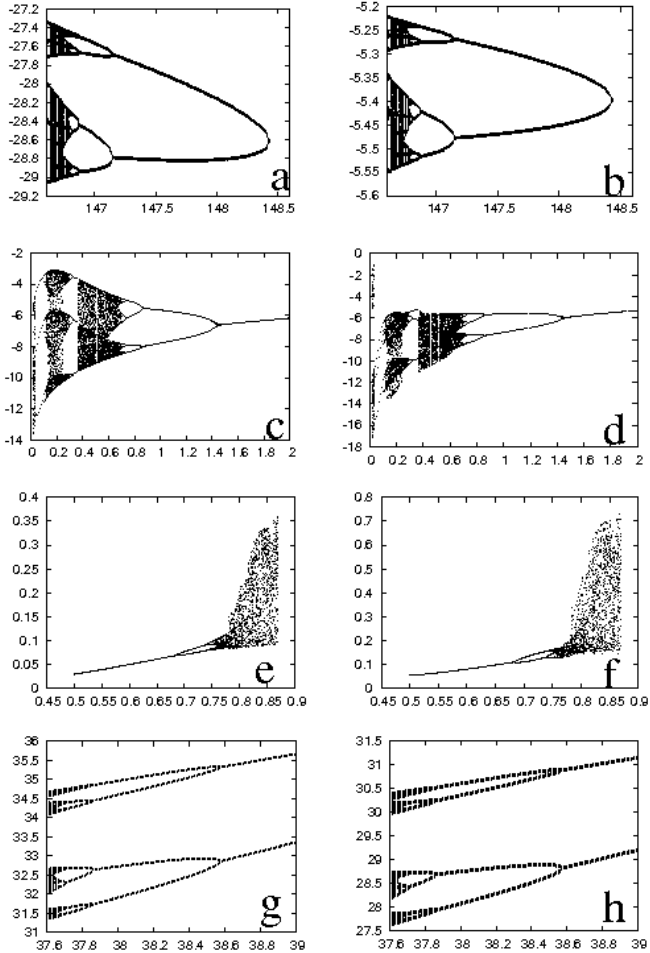}
\caption{Alikeness of BDs: figs. (a) and (b) show respectively the BDs of the driving and the driven Lorenz systems in complete replacement (CR) of $x_1$ coordinate; $r$ is varied for the driving and for driven it is kept fixed at $r=28$. $\sigma$ and $b$ are kept fixed at $10$ and $8/3$ respectively for the entire ESF. Similarly, in figs. (c) and (d) BDs are for R$\ddot{\textrm{o}}$ssler systems with unidirectional coupling in $x_1$ coordinate with coupling constant 1.6; $b$ of driving is varied. $a$ and $c$ are kept fixed at $0.2$ and $5.7$ respectively. Likewise, BDs of drivings and drivens of Chen system (figs. (e) and (f); CR in $x_2$) and Rabinovich-Fabrikant equation (figs. (g) and (h); CR in $x_2$) have been plotted. For all the figures, parameter is along the abscissa and variable is along the ordinate. For the standard forms of the equations one may refer \cite{9}.}
\end{figure}
\begin{figure}[h!]
\centering
\includegraphics[width=9.0cm]{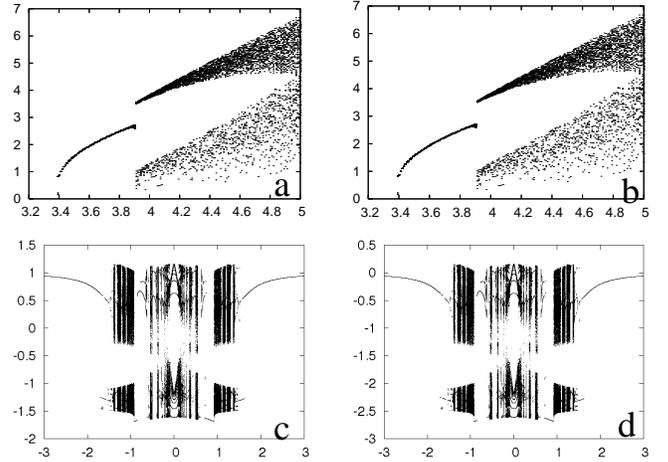}
\caption{Alikeness of BDs (continued): Two more examples of BDs of drivings and drivens of Rucklidge system (figs. (a) and (b); CR in $x_2$) and Duffing two well oscillator (figs. (c) and (d) CR in $x_2$). The striking alikeness is obvious. In the case of Rucklidge system the only parameter of the equation is varied. While, Duffing two well oscillator's BDs are w.r.t the variation of the angular velocity. For the standard forms of the equations one may refer \cite{9}.}
\end{figure}
The more important and interesting point to note is that one has, thus, induced bifurcation in $f_{driven}$ sans tampering with its parameter $p_0$.
This striking alikeness of BDs may be used as a very crude pictorial definition of extended synchronisation.
Also, it may serve to illustrate the robustness of GS, a property which requires to find a Lyapunov function, not an easy thing to do in general.
\\
\indent We now ask a deeper question: mathematically, when should two BDs be alike for the cases we are dealing with?
To answer this question let us first note that identical BDs would mean the existence of a family $D^P$ of $C^k$ diffeomorphism, each element of which is $d_{p,p_0}\ni D^P=\{d_{p,p_0}\ni  p\in P\}$ because the numerical experiments show that period changing in the driving system is causing same period changing in the driven system which, of course, hints that within an ESF (at least for those which have been experimented on numerically) not only GS but also conditional equivalence prevails (which can be checked by plotting the conjugate variables for an ESF for any parameter mismatch such that the parameters remain within $P$ and observing that there is no `kink' in the plot confirming at least $C^1$ diffeomorphism; an example will follow shortly).
Then, similar BDs would mean that for a particular Poincar$\acute{\textrm{e}}$ section in the driving system $\exists$ at least one such Poincar$\acute{\textrm{e}}$ section in the driven system $\ni$ $\exists$ a family, $M^P$ of homeomorphisms $m_{p,p_0} \ni M^P=\{m_{p,p_0}\ni  p\in P\}$ between the two Poincar$\acute{\textrm{e}}$ sections.
This introduces predictability in a broader sense: knowing BD of driving system one may predict the BD of driven system and the maps being bijective facilitates the existence of corresponding inverse maps allowing one to predict the BD of driving system from the knowledge of the BD of driven system w.r.t. parameter of the driving system.
\\
\indent As an aside, we would like to highlight a very interesting coincidence which may be pregnant with fruitful research possibilities.
In all the examples of ESF, on which numerical experiments have been done by us, the corresponding representative system may be written as a {\it jerk equation}\cite{9}.
(One may recall in the passing that in general a set of three coupled first order differential equation can't always be reduced to a jerk equation.)
{\it e.g.}, Lorenz: $\dddot{x}_1+(1+\sigma+b-\dot{x}_1/x_1)\ddot{x}_1+[b(1+\sigma+x_1^2)-(1+\sigma)\dot{x}_1/x_1]\dot{x}_1-b\sigma(r-1-x_1^2)x_1=0$,
R$\ddot{\textrm{o}}$ssler (with $a=b=\varepsilon$): $\dddot{x}_2+(c-\varepsilon+\varepsilon c-\dot{x}_2)\ddot{x}_2+[1-\varepsilon c-(1+\varepsilon^2)x_2+\varepsilon\dot{x}_2]\dot{x}_2+(\varepsilon x_2+c)x_2+\varepsilon=0$,
Rucklidge: $\dddot{x}_2+(1+\kappa-\dot{x}_2/x_2)\ddot{x}_2+\kappa(1-\dot{x}_2/x_2)+x_2^3=0$,
Nose-Hoover: $\dddot{x}_1-(\ddot{x}_1+x_1)\ddot{x}_1/\dot{x}_1-(a-1-\dot{x}_1^2)\dot{x}_1=0$ etc.
This means only one variable alongwith its derivatives is enough to carry all the information regarding the change in parameter being done to construct BD and unidirectional coupling of the driving system's ``single'' variable with the driven system is enough to induce the entire effect of the evolution of the driving system with the parameter into the driven system's BD w.r.t. the parameter of the driving system so that the two BDs are alike.
\section{BD IN R$\ddot{\textrm{O}}$SSLER-DRIVEN-LORENZ}
%
\begin{figure}[h!]
\centering
\includegraphics[width=9.0cm]{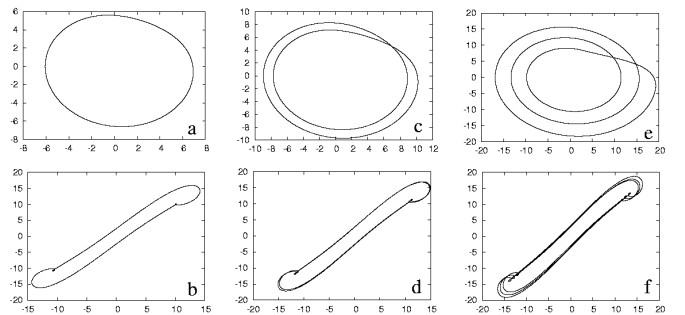}
\caption{Mimicry of R$\ddot{\textrm{o}}$ssler system's periodic multiplications by Lorenz system: figs. (a), (c) and (e) are of R$\ddot{\textrm{o}}$ssler system and corresponding figs. (b), (d) and (f) are of Lorenz system.}
\end{figure}
\begin{figure}[h!]
\centering
\includegraphics[width=9.0cm]{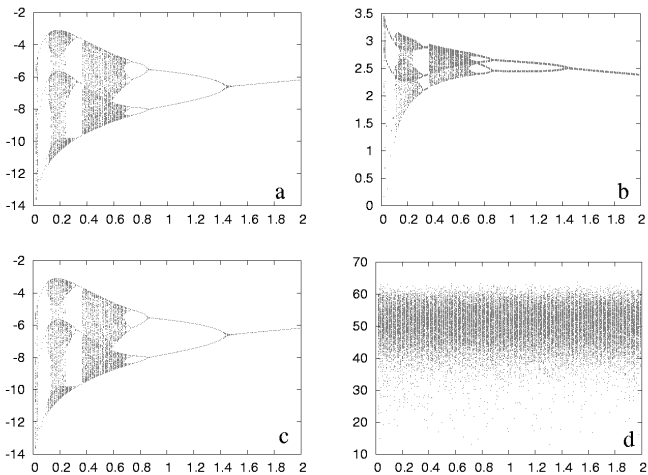}
\caption{BDs of R$\ddot{\textrm{o}}$ssler-driven-Lorenz: figs. (a) and (b) are for $k=40$ and are showing how BD for R$\ddot{\textrm{o}}$ssler (fig. (a)) is causing BD for Lorenz (fig. (b)) to be alike, while for $k=2$ one may see that BD for Lorenz (fig. (d)) is not following the BD for R$\ddot{\textrm{o}}$ssler (fig. (c)). $x_2$ and $y_2$ have been plotted w.r.t. $b4$.}
\end{figure}
\begin{figure}[h!]
\centering
\includegraphics[width=9.0cm]{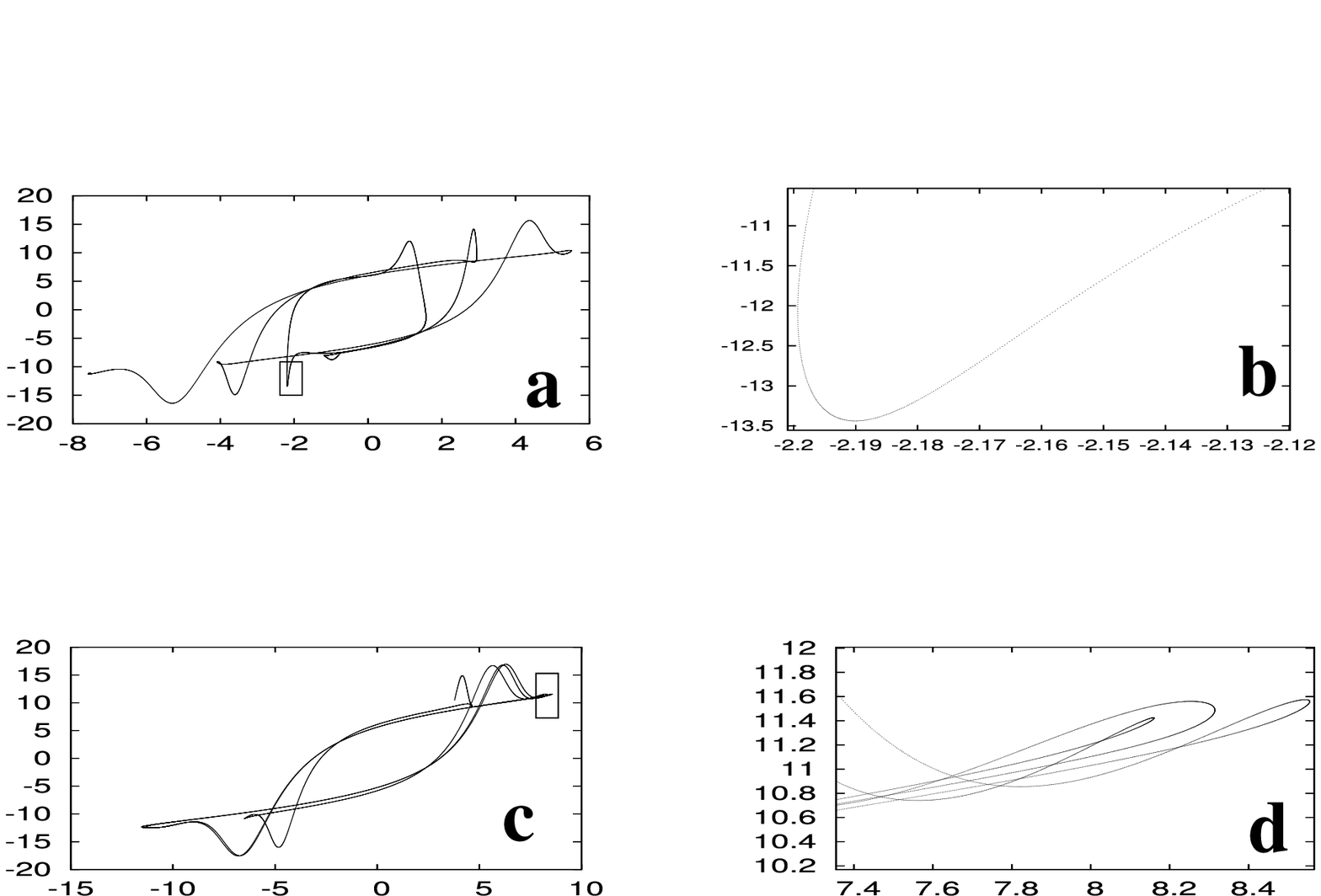}
\caption{Smoothness of the maps in R$\ddot{\textrm{o}}$ssler-driven-Lorenz for $k=40$: figs. (a) and (c) are respectively the plots for $y_2-x_2$ for the value $b=0.2$ and $0.75$ where there is chaos and period four respectively in the BD of R$\ddot{\textrm{o}}$ssler (fig. (5)). The figs. (b) and (d) are the blown up image of the boxed regions in the figs. (a) and (c) respectively, showing that there are no `kinks'. One should observe that the plot is not a line-plot but a dot-plot to appreciate the fact.}
\end{figure}
%
There remains one interesting possibility.
It can very well happen that given two ESFs, $F_i^{P_i}$ and $F_j^{P_j}$ the representative elements $f_i\in F_i^{P_i}$ and $f_j\in F_j^{P_j}$ respectively may depict GS when coupled.
For example, we can have a R$\ddot{\textrm{o}}$ssler system driving a Lorenz system through diffusive coupling with coupling constant $k$\cite{5}:
\begin{eqnarray}
\begin{array}{ll}
\textrm{R}\ddot{\textrm{o}}\textrm{ssler}& \phantom{x}\textrm{Lorenz (diffusively coupled)}\\
\dot{x_1}=-(x_2+x_3) & \phantom{x}\dot{y_1}=\sigma(y_2-y_1)\\
\dot{x_2}=x_1+ax_2& \phantom{x}\dot{y_2}=ry_1-y_2-y_1y_3+k(x_2-y_2)\\
\dot{x_3}=b+x_3(x_1-c)& \phantom{x}\dot{y_3}=y_1y_2-by_3
\end{array}
\label{3}
\end{eqnarray}
This arrangement is known to show GS for $a=b=0.2,c=9.0,\sigma=10,r=60,b=8/3$ and most importantly $k=40$.
In the passing, it may be noted that the driving system may again be written as a jerk equation as mentioned earlier.
To capture this GS, one may painstakingly go on to calculate the continuity and differentiability statistics\cite{10} developed by Pecora {\it et. al.}
Of course, the process is rewarding with computers making the job easy.
With the parameter values fixed at the values mentioned above, one may note (as in fig.(4)) how the period doubling and the period tripling in R$\ddot{\textrm{o}}$ssler system brought about with the variation in $c$ is mimicked by the Lorenz system whose parameters are kept fixed.
Also, the time series of the driving system in non-chaotic state stops the chaos in the driven just as the time series of the driving system in chaotic state can initiate the chaos in the driven even if it initially was in non-chaotic state.
This mimicry is gradually lost when $k$ is reduced as is evident from the fig.(5).
Hence, even in this case BDs have been able to showcase the existence GS when it really exists and this is in consistence with the statistical technique used by Pecora {\it et. al.}
As with the GS between the elements of a single ESF, the existence of conditional equivalence may also be claimed here because the maps for various values of $b$ are smooth.
At typical values of $b$ (fig. (6)), one may note that there are no kinks in the plot ensuring at least $C^1$ map.
\section{CONCLUSION}
To conclude, we emphasis that the numerical study made in this paper encourages to take up further research on the apparent connection between BD and GS by divulging the facts how BDs can assist in assessing the robustness of GS and predictability of nature of the systems in GS.
The classification of various chaotic systems in equivalence classes have been done.
BD has been shown to be a distinguishing characteristic of such an equivalent class.
In future, similar studies on chaotic maps and bi-directionally coupled synchronised chaotic systems would be of interest.
\\
\\
\indent CSIR (India) is gratefully acknowledged for awarding fellowship to SC, one of the authors. Also, Prof. J.K. Bhattacharjee and  Prof. R. Ramaswamy are gratefully acknowledged for helpful discussions.


\begin{thebibliography}{99}
%
\bibitem{1a}T. Yamada and H. Fujisaka, {Prog. Theor. Phys. {\bf {70}}}, 1240(1983)
%
\bibitem{1b}T. Yamada and H. Fujisaka, {Prog. Theor. Phys. {\bf {72}}}, 885(1984)
%
\bibitem{2}V. S. Afraimovich, N. N. Verichev, and M. I. Rabinovich, {Inv. VUZ Rasiofiz. RPQAEC {\bf{29}}}, 795(1986)
%
\bibitem{3}L. M. Pecora and T. L. Carroll, {Phys. Rev. Lett. {\bf{64}}}, 821(1990)
%
\bibitem{4} E. Ott, C. Grebogi, and J. A. Yorke, {Phys. Rev. Lett. {\bf{64}}}, 1196(1990)
%
\bibitem{5}L.M. Pecora, T. L. Carroll, G.A. Johnson, D.J. Mar, and J.F. Heagy, {Chaos {\bf{7}}(4)}, 520(1997)
%
\bibitem{6}N.F. Rulkov, M.M. Sushchik, and L.S. Tsimring, {Phys. Rev. E {\bf{51}}}, 980(1995)
%
\bibitem{7}L. Kocarev and U. Parlitz, {Phys. Rev. Lett. {\bf{76}}}, 1816(1996) 
%
\bibitem{8}M. Nakahara, {\it Geometry, Topology and Physics, Second Edition}, (IOP Publishing Ltd)
%
\bibitem{9}J.C. Sprott, {\it Chaos and Time-Series Analysis}, (Oxford University Press)
%
\bibitem{10}L.M. Pecora, T. L. Carroll, and J.F. Heagy, {Phys. Rev. E {\bf{52}}}, 3420(1995)
%
\end{thebibliography}
\end{document}